\documentclass[aps,preprint,onecolumn,eqsecnum]{revtex4}
\usepackage{dcolumn}
\usepackage{bm}
\usepackage{amssymb}
\usepackage{amsmath}
\usepackage{amsfonts}

\usepackage[]{graphicx}
\begin{document}
\bibliographystyle{prsty}
\title{Brownian motion in the pilot wave interpretation of de Broglie and relaxation to quantum equilibrium }
\author{ Aur\'elien Drezet $^{1}$}
\address{(1) Univ. Grenoble Alpes, CNRS, Institut N\'{e}el, F-38000 Grenoble, France
}
\begin{abstract}
 The pilot wave interpretation proposed by de Broglie and later by Bohm contains not only a dynamical ontology but also relies on a statistical assumption known as quantum equilibrium. In this  work which follows our recent article \cite{Drezet} we develop a Langevin force description of the relaxation process which leads to quantum equilibrium.   Based on a application of the Caldera-Leggett model for a thermal bath we show how a Brownian motion leads naturally to quantum relaxation.   
\end{abstract}

 \maketitle
\section{Introduction}
\indent In a recent article published in this journal~\cite{Drezet} we discussed the issue of how to justify the so called `Born's rule' for quantum probability in the context of the `hidden-variable' theory proposed by de Broglie~\footnote{De Broglie's pilot wave was a consequence of the double solution theory he proposed~\cite{deBroglie1927,livre}. While this topic is of fundamental importance (much more than the subject of the present article) we will not discuss it here. For more on this subject see for example the special issues `\textit{Quantum Rogue Waves as Emerging Quantum Events}' in Ann. Fond. L. de Broglie \textbf{42} and specially the reviews~\cite{Fargue,Durt}).} in 1925-1927~\cite{deBroglie1927,livre} later rediscovered by Rosen in 1942~\cite{Rosen1942} and Bohm in 1952~\cite{Bohm1952} and known as the pilot wave interpretation (PWI) or recently `Bohmian mechanics'. After reviewing several important  proposals for solving this issue we advocated a stochastic approach based on a Fokker-Planck or diffusion equation reminiscent of studies about the classical Brownian motion.\\ 
\indent More specifically, based on the seminal work by Bohm and Vigier in 1954~\cite{BohmVigier1954} and  Vigier in 1956~\cite{Vigier1956} we derived a diffusion-like equation for the density of probability $\rho(x,t)$ for finding a quantum particle at spatial location $x$ and time $t$ when the system is coupled to a thermostat. We showed that on the long term $\rho(x,t)$ necessarily converges to the usual quantum prediction $\rho_\psi(x,t)=|\psi(x,t)|^2$ where $\psi(x,t)$ is the Schrodinger wave function associated with the particle. We also connected our work to Boltzmann's  derivation of the second law of  thermodynamics and derived a quantum version of the H-theorem  $dH_t/dt\leq 0$ (different of the `Gibbs-Tolman' coarse-graining proposed by Valentini in 1991~\cite{Valentini1991,Valentini2005}) and which demonstrates the irreversible tendency to reach quantum  equilibrium  $\rho_\psi$ within the condition of application of our model. We emphasize that our approach like the one of Valentini are not necessarily orthogonal to the typicality interpretation advocated  by D\"{u}rr, Goldstein and Zanghi~\cite{Durr}. In all these approaches we exploit some results obtained by Boltzmann in thermo-statistics and in kinetic theory. Indeed, some notions of typicality must be included as well in the discussion of the H-theorem and our aim with diffusion was mainly to show that the dynamics is robust enough for going beyond a simple statement of typicality (associated with a simple `branch'-counting process: see \cite{Drezet} for a discussion). At the end of the article we emphasized the key role of entanglement and decoherence with the environment. We believe that these features associated with deterministic chaos can be used to enlarge the conditions of typicality developed in \cite{Durr}. \\
\indent In this context and very recently, during an interesting conference on Quantum Foundations at Troyes-France we were asked~\cite{conference} how to define a numerical estimation of the diffusion constant $D$ appearing in our model. Indeed, in our approach~\cite{Drezet} the nature of  the interaction process between the particle and the thermostat was not discussed in details. This is however a fundamental issue and here we provide an elementary theory for defining the diffusion constant $D$. For this purpose we will introduce a PWI version of the Langevin equation for quantum Brownian motion. In our model based on the standard Caldeira-Leggett approach~\cite{Caldeira} for coupling a particle to a bath of harmonic oscillators we will be able to define  a PWI version of the  generalized Langevin equation including a quantum potential  \`a la de Broglie-Bohm. 
Our approach is only based on the deterministic PWI framework and can be understood as an attempt to include some elements of decoherence and Langevin-Noise theory in the ontology of de Broglie-Bohm. Since this ontology is fundamentally nonlocal and holistic this issue is not trivial as we will show in this manuscript. Importantly, since we stick with determinism our approach differs from the stochastic models developed for example by Nelson or for stochastic  quantum electrodynamics (SQED). More precisely, in the discussion we will have to consider the role of the so called Schrodinger-Langevin equation proposed by Kostin in 1972~\cite{Kostin1972,Sanz2014}. This will be the occasion to go back to  some earlier proposals by Bohm and Hiley~\cite{Hiley}, Furth, Fenyes, Nelson or Luis de la Pe\~{n}a~\cite{Furth,Fenyes,Nelson,Bacciagaluppi,Luis,Cushing}, and  de Broglie himself~\cite{Jalons} based on a `subquantum dynamics'~\cite{BohmVigier1954} or an `hidden thermodynamics'.  We emphasize that these earlier proposals essentially relied on a yet unknown  level of reality - far below the existing quantum level- and associated with some `subplanckian' stochastic fluctuations in a hypothetical `Dirac Aether' advocated by Vigier and Bohm or Nelson. In these approaches the irregular motions of such a complex background fluid would generate a Brownian motion for the quantum particle. Our model has a much less ambitious goal and actually relies strictly on the firm basis of current and accepted quantum mechanics, i.e., on the Schrodinger equation and on the quantum theory of open systems applied to the PWI. This has a huge consequence because the relaxation mechanism provided by our theory has only a meaning when the quantum system considered is interacting with a thermostat associated with a bath of oscillators (all of these quantum objects obeying to a single complex Schrodinger equation in agreement with the philosophy of the PWI). Therefore, in our approach, at the difference of the earlier proposals quoted before which involved a subquantum level, there is no  anymore relaxation for free particles such as electrons or atoms after being emitted from a (thermal) source. However, since the Liouville theorem preserves the quantum equilibrium once it is (approximately) reached the Born rule  $\rho_\psi(x,t)=|\psi(x,t)|^2$ will be always experimentally verified with a high accuracy for any quantum object well prepared and separated from a source in which quantum relaxation already occurred due to thermal interaction. Since this relaxation will be very fast the probability to find a disagreement with the standard quantum prediction will thus be always vanishingly small. Of course PWI opens new gates since  the Born rule is not imposed as a statement (unlike in the conservative Copenhagen approach). Therefore, deviations to quantum equilibrium are always possible at least in the early ages of the Universe~\cite{ValentiniPRD} where equilibrium is not yet reached or where the particle wavelength is larger than the instantaneous Hubble radius. This could induces violation of the no-signaling theorem prohibiting effective faster than light communications~\cite{Valentini1991}. It would be of paramount importance to search seriously some residual relics or signatures of this non-locality and quantum non-equilibrium in the cosmological background. These important issues and many related ones will  however not be considered here. 
\section{The quantum Brownian motion seen from the perspective of the pilot wave interpretation }                       
\indent We start with a rapid description of the classical version of the Caldeira-Legget model~\cite{Caldeira} for a particle  $S$ of mass $m$ in a external potential $V(x)$ and coupled to a bath  T of harmonic oscillators. The Hamiltonian for this system is given by
\begin{eqnarray}
H=\frac{p^2}{2m}+V(x)+\sum_n \frac{p_n^2}{2m_n}+\frac{m_n\omega_n^2}{2}(x_n-\frac{c_n x}{m_n\omega_n^2})^2\label{1}
\end{eqnarray} where $p$ is the canonical momentum conjugated to the coordinate $x$ for the subsystem $S$ while  $x_n$ and $p_n$ are canonical variables for the various oscillators  of mass $m_n$ and pulsation $\omega_n$ of the reservoir  T (labeled by $n$). In the model there is a coupling constant $c_n$ between the particles of S  and T. The structure of this model is well documented in the literature: it was proposed by Ford, Kac an Mazur in 1965 \cite{Ford1965} but it was popularized after the work by Caldeira and Leggett~\cite{Caldeira} (for a complete discussion see for example \cite{livrestat}). Based on the Hamilton equations and Eq.~\ref{1} we derive easily the set of coupled Newton's equations describing the complete dynamics:
\begin{eqnarray}
m\ddot{x}=-\nabla V(x)+\sum_n c_n(x_n-\frac{c_n x}{m_n\omega_n^2})	\label{2}
\end{eqnarray}
\begin{eqnarray}
m_n(\ddot{x}_n+\omega_n^2x_n)=c_nx	\label{3}
\end{eqnarray}
 Before solving this system it is useful to go directly to the PWI to see how the equations will be modified. 
In the the PWI the fundamental equation is the Schrodinger equation $i\hbar\partial\psi_t =\hat{H}\psi_t$ for the the full system where $\hat{H}$ is now an Hermitian Hamilton operator. The Standard procedure for defining a quantum version of the Caldeira-Leggett model is thus to go to the Heisenberg representation and to solve like in classical physics the set of Eqs.~\ref{2}, and \ref{3}. However, in the PWI the most useful representation is the Madelung-de Broglie one which relies on the nonlinear polar expression $\psi_t=a_t e^{iS_t/\hbar}$ where $a$ and $S$ are respectively the amplitude and phase of the wave function. Since we work in the configuration space we have $a_t=a(x(t),\{x_n(t)\},t)$, $S_t=S(x(t),\{x_n(t)\},t)$.  With the guidance law $m_n\dot{x}_n(t)=\nabla_n S_t=p_n$ and $m\dot{x}(t)=\nabla S_t=p$   we obtain the well-known Hamilton Jacobi equation
\begin{eqnarray}
-\partial_t S(x,\{x_n\},t)=H(x,p,\{x_n,p_n\})+Q(x,\{x_n\},t)
\end{eqnarray} where $H(x,p,\{x_n,p_n\})$ is the classical Hamiltonian given in Eq.~\ref{1} and $Q(x,\{x_n\},t)$  is the in general  highly non-local quantum potential introduced by de Broglie and which reads here \begin{eqnarray}
Q(x,\{x_n\},t)= \frac{-\hbar^2}{2m}\frac{\nabla^2 a(x,\{x_n\},t)}{a(x,\{x_n\},t)}+ \sum_n\frac{-\hbar^2}{2m_n}\frac{\nabla_n^2 a(x,\{x_n\},t)}{a(x,\{x_n\},t)}
\end{eqnarray}
Now, from the Hamilton Jacobi Equation we can easily rederive the Newton equations like in Eq.~\ref{2} and \ref{3} but this time with the new Hamiltonian  $H(x,p,\{x_n,p_n\})+Q(x,\{x_n\},t)$. This leads directly to  
\begin{eqnarray}
m\ddot{x}=-\nabla(V(x)+Q(x,\{x_n\},t))+\sum_n c_n(x_n-\frac{c_n x}{m_n\omega_n^2})	\label{2b}
\end{eqnarray}
\begin{eqnarray}
m_n(\ddot{x}_n+\omega_n^2x_n)=c_nx	-\nabla_n Q(x,\{x_n\},t)\label{3b}
\end{eqnarray} which differ from the previous set by the inclusion of the (nonlocal) quantum forces $-\nabla Q(x,\{x_n\},t)$ and $-\nabla_n Q(x,\{x_n\},t)$.\\
\indent  At that stage we mention briefly, as it was already pointed out by   Takabayasi  in 1953 \cite{Takabayasi}, that  the PWI   written in the Newton form must be supplied with the guidance condition $m_n\dot{x}_n(t)=\nabla_n S_t=p_n$ and $m\dot{x}(t)=\nabla S_t=p$ which imposes to the velocity to be the gradient of a phase. Schrodinger's equation also imposing the single-valuedness of the wave function at any point in the configuration space, the phase $S$ may have some discontinuities since around any closed loop of this space the quantization condition $\oint \nabla S dx= 2\pi n$ (with $n$ an integer and $n\neq 0$ is associated with vortex lines) holds. This condition stirred some controversies about the equivalence between the first order and second order dynamics~\cite{Wallstrom,Holland,colin}. Here we will not enter into this debate and assume that  Eqs.~\ref{2b} and  \ref{3b} also satisfy the single-valuedness constraints (for this purpose it is enough to consider that $p$ and $p_n$ are obeying the guidance law, i.e., defined as a phase gradient at a given time $t_0$ which could be the origin).\\ 
\indent Now we go back to the integration of the dynamical equations. From   Eq.~\ref{3b} we directly get the formal solution
\begin{eqnarray}
x_n(t)= x_n^{(0)}(t)+\int_{t_0}^{t}dt'\frac{\sin{(\omega_n(t-t'))}}{m_n\omega_n}[c_nx(t')	-\nabla_n Q(x(t'),\{x_n(t')\},t')]\nonumber\\
\label{3d}
\end{eqnarray}
where $x_n^{(0)}(t)=x_n(t_0)\cos{(\omega_n(t-t_0))}+\frac{\dot{x}_n(t_0)}{\omega_n}\sin{(\omega_n(t-t_0))}$ is the general  free solution defined with the boundary conditions at time $t_0$.  We emphasize that $x_n^{(0)}(t)$ is a  classical-like solution of $m_n(\ddot{x}_n+\omega_n^2x_n)=0$, i.e., when there is no interaction and no quantum potential. Therefore, the physical meaning of $x_n^{(0)}(t)$ is not automatic in the PWI where quantum forces $\nabla_n Q$ in general never vanish and depends of the quantum states $\psi$ chosen. This issue will become important later.  From now, inserting Eq.~\ref{3d} into Eq.~\ref{2b} leads to the generalized Langevin  equation for $x(t)$:
 \begin{eqnarray}
m\ddot{x}(t)=-\nabla(V(x(t))+Q(x(t),\{x_n(t)\},t))\nonumber\\
-m\gamma(t-t_0)x(t_0)-m\int_{t_0}^{t}dt'\gamma(t-t')\dot{x}(t')+F(t)+\Delta	\label{2c}
\end{eqnarray}
 in which the memory friction reads 
 \begin{eqnarray}
\gamma(\tau)=\frac{1}{m}\sum_n\frac{c_n^2}{m_n\omega_n^2}\cos{(\omega_n \tau)}\label{perte}
\end{eqnarray} 
and the fluctuating force is 
  \begin{eqnarray}
F(t)=\sum_n c_n x_n^{(0)}(t).\label{force}
\end{eqnarray} 
Importantly, Eqs.~\ref{perte} and \ref{force} are identical in the quantum and classical case, i.e. if we neglect the quantum forces. The specific terms arising from the PWI are the nonlocal gradient $-\nabla Q(x(t),\{x_n(t)\},t)$  and the nonlocal force $\Delta$ which reads 
 \begin{eqnarray}
\Delta=-\int_{t_0}^{t}dt'\sum_n\frac{c_n}{m_n\omega_n}\sin{(\omega_n (t-t'))}\nabla_n Q(x(t'),\{x_n(t')\},t')
\label{quantum}
\end{eqnarray} In Eq.~\ref{quantum} the nonlocality is even double since it appears in the quantum potential (we thus speak of nonlocality \`a la Bell) and in the time integral (this second kind of nonlocality in time is associated with memory effects or hereditary dynamics and has a more classical origin going back at least to V. Volterra and L. Boltzmann).\\
\indent The present model is quite general but its level of complexity is such that in order to get a practical solution we must add some hypotheses to simplify the description.  For this purpose we go back to our previous paper \cite{Drezet} and point out that at some stage in the derivation of the diffusion equation we admitted a factorization \textit{ansatz} $\rho_{S+T}(x,\{x_n\},t)\simeq\rho_S(x,t)\rho_T(\{x_n\},t)$  and  $|\psi_{S+T}(x,\{x_n\},t)|^2\simeq|\psi_S(x,t)|^2|\psi_T(\{x_n\},t)|^2$ near the equilibrium. This axiom is reminiscent of the old  `molecular chaos' introduced by Boltzmann and it also appears under the name of Born-Markov  approximation in the context of  quantum-like master equations~\cite{cohen}. This is often used in the literature  together with system-reduced density matrix calculations such as it is done within the Redfield or Lindblad approaches. Actually, we see that here this hypothesis implies the amplitude relation $a_{S+T}(x,\{x_n\},t)\simeq a_S(x,t)a_T(\{x_n\},t)$  but that the phase is not impacted by the reasoning so that we still keep the entanglement complexity in $ S(x,\{x_n\},t)$. Moreover, from the amplitude factorization we deduce $Q(x,\{x_n\},t)=Q_S(x,t) +Q_T(\{x_n\},t)$ with 
\begin{eqnarray}
Q_S(x,\{x_n\},t)= \frac{-\hbar^2}{2m}\frac{\nabla^2 a_S(x,t)}{a_S(x,t)}\nonumber\\
Q_T(\{x_n\},t)= \sum_n\frac{-\hbar^2}{2m_n}\frac{\nabla_n^2 a_T(\{x_n\},t)}{a_T(\{x_n\},t)}.\label{postu}
\end{eqnarray}  Therefore Eq.~\ref{2c} now reads
\begin{eqnarray}
m\ddot{x}(t)=-\nabla(V(x(t))+Q_S(x(t),t))\nonumber\\
-m\gamma(t-t_0)x(t_0)-m\int_{t_0}^{t}dt'\gamma(t-t')\dot{x}(t')+F(t)+\Delta	\label{2d}
\end{eqnarray} 
with   \begin{eqnarray}
\Delta=-\int_{t_0}^{t}dt'\sum_n\frac{c_n}{m_n\omega_n}\sin{(\omega_n (t-t'))}\nabla_n Q_T(\{x_n(t')\},t').
\label{quantumb}
\end{eqnarray} The advantage of this new dynamics is that the motion of  S and T can be in principle solved. However,  the model is still too complex for the present purpose. Ideally, we would like to remove or neglect the effect of the quantum potential  $Q_T(\{x_n(t')\},t')$. This would be apparently justified if the temperature of the bath is high so that the motions $x_n(t)$ are supposed to be  quasi-classical. However the meaning of quasi-classical states of the environment is ambiguous in the PWI. For example the usual semi-classical WKB states of the harmonic oscillator  have some pathological features. Indeed, it is well known that in such stationary WKB states the guidance velocity $\nabla_n S^{(0)}_n/m_n$ of the non interacting harmonic oscillators vanishes and the associated quantum potential $Q_T^{(0)}$ survives~\cite{Hiley}. Therefore, these states are from the point of view of the PWI highly non classical since there is no kinetic energy and the role of the quantum potential becomes dominant (we point out that Einstein and Rosen dismissed the PWI because of this difficulty).  Here, instead of the WKB states we should better consider the coherent (or Gaussian) states which naturally emerge as the only privileged states through decoherence (i.e., continuous monitoring) resulting from interactions with `the rest of the universe'~\cite{zurek,Appleby}. Importantly, the coherent states are characterized by classical trajectories, i.e., up to an additional restoring force term (see the discussion in Appendix) due to a residual quantum potential contribution.\\
\indent  In order to use these states in our problem we return to Eq.~\ref{3d} and we write instead:
\begin{eqnarray}
x_n(t)= x_n^{(\alpha_n)}(t)+\int_{t_0}^{t}dt'\frac{\sin{(\omega_n(t-t'))}}{m_n\omega_n}[c_nx(t')\nonumber\\	-\nabla_n Q'(x(t'),\{x_n(t')\},t')]
\label{3e}
\end{eqnarray}
where $x_n^{(\alpha_n)}(t)$ is the bohmian trajectory of the $n^{th}$ oscillator if this system is characterized  by the coherent state $\psi_n^{(\alpha_n)}(x_n,t)$ corresponding to  the complex number $\alpha_n(t)$ (see Appendix) and the boundary condition $x_n^{(\alpha_n)}(t_0)=x_n(t_0)$. We have:
 \begin{eqnarray}
x_n^{(\alpha_n)}(t)=\sqrt{\frac{2\hbar}{m_n\omega_n}}|\alpha_n(t_0)|\cos{(\omega_n (t-t_0)-\sigma_n)}+u_n\nonumber\\
=x_n^{(0)}(t)-\int_{t_0}^{t}dt'\frac{\sin{(\omega_n(t-t'))}}{m_n\omega_n}\nabla_n Q_n^{(\alpha_n)}(x_n^{(\alpha_n)}(t'),t')
\label{3f}
\end{eqnarray} where $u_n$ and $\sigma_n$ are constants defined in the Appendix (see Eq.~\ref{sol2}). The quantum potential $Q_n^{(\alpha_n)}$ is defined in Eq.~\ref{quantumpot} and for consistency the new quantum potential $Q'$ in Eq.~\ref{3e} is defined as $Q'(x(t),\{x_n(t)\},t)=Q(x(t),\{x_n(t)\},t)-\sum_n Q_n^{(\alpha_n)}(x_n^{(\alpha_n)}(t),t)$. With Eq.~\ref{3e} we can replace Eq.~\ref{2c} by 
\begin{eqnarray}
m\ddot{x}(t)=-\nabla(V(x(t))+Q(x(t),\{x_n(t)\},t))\nonumber\\
-m\gamma(t-t_0)x(t_0)-m\int_{t_0}^{t}dt'\gamma(t-t')\dot{x}(t')+F'(t)+\Delta'	\label{2new}
\end{eqnarray}
 in which the memory friction is left unchanged and where the new fluctuating force is 
  \begin{eqnarray}
F'(t)=\sum_n c_n x_n^{(\alpha_n)}(t),\label{forcenew}
\end{eqnarray} 
while the nonlocal force becomes 
 \begin{eqnarray}
\Delta'=-\int_{t_0}^{t}dt'\sum_n\frac{c_n}{m_n\omega_n}\sin{(\omega_n (t-t'))}\nabla_n Q'(x(t'),\{x_n(t')\},t').
\label{quantumnew}
\end{eqnarray}
This new description is rather formal until we go back to Eq.~\ref{postu}. Here, as explained in the Appendix, we should consider for the thermostat a mixture of coherent states $\alpha_n$ (more precisely a mixture of product states $\otimes_n|\alpha_n\rangle$). In the PWI,  where there is only one wave function for the whole universe, this actually means that due to interaction with the rest of the universe the density matrix of the bath T is well approximated by such a mixture. Therefore, if mathematically we isolate one of this product state $\otimes_n|\alpha_n\rangle$ and apply the Born-Markov approximation starting from time $t_0$ where S and T are decoupled it is reasonable to write  $Q_T(\{x_n(t)\},t)\simeq \sum_n\frac{-\hbar^2}{2m_n}\frac{\nabla_n^2 a_n^{(\alpha_n)}(x_n(t),t)}{a_n^{(\alpha_n)}(x_n(t),t)}\simeq \sum_n Q_n^{(\alpha_n)}(x_n^{(\alpha_n)}(t),t)$. This assumes that the trajectories of the bath are weakly affected by the interaction with S and that the amplitudes $a_n^{(\alpha_n)}$ (and thus the quantum potential of the bath) are not modified.\\
\indent  Within this approximation  the nonlocal force $\Delta'$ vanishes and we have finally 
 \begin{eqnarray}
m\ddot{x}(t)\simeq-\nabla(V(x(t))+Q_S(x(t),t))\nonumber\\
-m\gamma(t-t_0)x(t_0)-m\int_{t_0}^{t}dt'\gamma(t-t')\dot{x}(t')+F'(t).	\label{2truc}
\end{eqnarray} 
Under this form we have the generalized Langevin equation with retardation and colored noise force $F'(t)$ in presence of the external potential  $V(x)$ and of the effective quantum potential $Q_S(x,t)$. Like in classical physics we would like to write the fluctuation-dissipation theorem of the second kind assuming a thermal classical bath:
  \begin{eqnarray}
	\langle F'(t)\rangle_{th.}=0,& &
C_F(\tau)=\langle F'(t_b)F'(t_a)\rangle_{th.}=m k_B T\gamma(t_b-t_a).\label{therm}
\end{eqnarray} The meaning of the averaging $\langle[...]\rangle_{th.}$ over thermal bath degrees of freedom is standard in classical physics but should be clarified  a little in the context of quantum mechanics and PWI  where the primary reality is the universal wave function associated with a pure quantum state (while a thermal state is a mixture). This issue is discussed in the Appendix.
In the PWI model we can thus easily demonstrate $\langle F'(t)\rangle_{th.}=0$ implying that like in classical physics the net random force vanishes. The two times force correlator $C_F^{(PWI)}(\tau)=\langle F'(t_b)F'(t_a)\rangle_{th.}^{(PWI)}$ is more difficult to define and to calculate. The details are given in the Appendix and the superscript PWI here indicates that the meaning of the force product is taken in the PWI sense not in the usual operator sense. We find explicitly 
  \begin{eqnarray}
	C_F^{(PWI)}(\tau)=A+k_BTm\gamma(\tau).
\label{thermopwi}
\end{eqnarray} The additional  contribution $A=\sum_n \frac{c_n^2}{m_n\omega_n^2}\frac{\hbar\omega_n}{2}$ is specific of the quantum model considered and is a signature of a zero point field (zpf) fluctuation in the PWI. We have the general  constraint $A\ll k_BTm\gamma(0)$ which can be deduced from the condition $\hbar\omega_n\ll k_BT$. We emphasize that $A$ contrarily to $\gamma(\tau)$ is not decaying with time (i.e., $\texttt{lim}_{\tau\rightarrow+\infty}[\gamma(\tau)]=0$). Therefore $A$ is associated with a form on nonlocality and correlation in time specific of the PWI. To precise the meaning of $A$ we can use the continuous limit and write $\gamma(\tau)=\int_0^{\omega_c} d\omega g(\omega)\cos{(\omega\tau)}$ and $A=m\int_0^{\omega_c} d\omega g(\omega)\frac{\hbar\omega}{2}$ where $\omega_c^{-1}$ defines a memory time scale  and $\hbar\omega_c\ll k_BT$. We here limit our study to  non-retarded dissipation regime  with infinitely short memory time~\footnote{ We consider the simple Ohmic model where $\gamma(\tau)=\omega_c \Gamma e^{-\omega_c|\tau|}$ for which   in the limit $\omega_c\rightarrow +\infty$ we have $\gamma(\tau)=2\Gamma\delta(\tau)$. This limit allows us to recover  Eq.~\ref{2f}. } $g(\omega)\simeq \frac{2}{\pi}\Gamma$, $\gamma(\tau)\simeq 2\Gamma\delta(\tau)$ associated with a white noise. It leads to the Markovian-Langevin equation
\begin{eqnarray}
m\ddot{x}(t)\simeq-\nabla(V(x(t))+Q_S(x(t),t))
-m\Gamma\dot{x}(t)+F'(t).	\label{2f}
\end{eqnarray}  This model is very close to the classical case and the main differences come from the presence of a quantum potential contribution $Q_S(x,t)$ and the inclusion of the constant $A$. In this model we can fairly write $A\simeq \frac{\Gamma m\hbar \omega_c^2}{2\pi}$ which shows how $A$ typically depends on $\omega_c$ and $\Gamma$. The presence of $A$ leads therefore to unusual features and the influence would become important a low temperature (a regime not considered here for questions of space). $A$ is connected to the fundamental fluctuation in force $\langle F'^2\rangle^{(Quantum)}=\Delta F'^2=\sum c_n^2 \Delta x_n^2$ where $\Delta x_n=\sqrt{(\frac{\hbar}{2m_n\omega_n})}$ is the fundamental uncertainty associated with the Gaussian wave packet of the $n^{th}$ bath  oscillator. It is thus intrinsically quantum and  from the procedure defined here it is the minimal fluctuation available so that further approximations would only make this fluctuation worst and induce even more nonlocality. \\
\indent Now, for illustration we can locally take in Eq.~\ref{2f}
$V(x)=constant$ and $a_S(x)$ will be also  spatially uniform meaning that the average motion is a plane wave in a constant potential. Such a situation will be a good approximation in rarefied medium like molecular gases or for free electrons in solids in the Drude approximation. Then the force $-\nabla(V(x(t))+Q_S(x(t),t))$ approximately vanishes and we obtain a form of Brownian motion such that in the limit $\Gamma(t-t_0)\gg 1$, $\Gamma \tau \gg 1$:
\begin{eqnarray}
\langle (\dot{x}(t))^2\rangle_{th}\simeq \frac{k_B T }{m}+\frac{A}{m^2\Gamma^2}=\frac{k_B T }{m}(1+\frac{\hbar \omega_c}{k_BT}\frac{\omega_c}{2\pi\Gamma})\nonumber\\
\langle |x(t+\tau)-x(t)|^2\rangle_{th}\simeq \frac{2 k_B T }{m\Gamma} \tau + \frac{A}{m^2\Gamma^2}\tau^2.\label{einstein}\end{eqnarray}
 The first line is in agreement with the equipartition theorem if we introduce an effective temperature $T_{eff}=T(1+\frac{\hbar \omega_c}{k_BT}\frac{\omega_c}{2\pi\Gamma})$. This effective temperature is in general  different of $T$. Indeed while we are in the limit $\frac{\hbar \omega_c}{k_BT}\ll 1$  we have also often (but not always see below)  $\frac{\omega_c}{2\pi\Gamma}\gg 1 $ so that the two ratios  generally compete.  The second line in Eq.~\ref{einstein} allows us to  define the diffusion `constant' as 
\begin{eqnarray}
D=\frac{\langle |x(t+\tau)-x(t)|^2\rangle_{th}}{2\tau}=\frac{k_B T}{m\Gamma}+ \frac{A}{2m^2\Gamma^2}\tau\nonumber\\
=\frac{k_B T }{m\Gamma}(1+\frac{\hbar \omega_c}{k_BT}\frac{\omega_c}{2\pi\Gamma}\frac{\Gamma\tau}{2})=\frac{k_B T }{m\Gamma}(1+\frac{\hbar \omega_c}{k_BT}\frac{\omega_c\tau}{4\pi}).\label{diff}
\end{eqnarray}We have clearly two regimes: a pure diffusive, i.e. Einsteinian, one $$\sqrt{(\langle |x(t+\tau)-x(t)|^2\rangle_{th})}\simeq \sqrt{\frac{2 k_B T }{m\Gamma}} \sqrt{\tau}$$ if $\hbar\omega_c\ll \sqrt{(k_BT\hbar/\tau)}\ll\sqrt{(k_BT\hbar\Gamma)}$ (i.e., $T\simeq T_{eff}$) and a linear spreading regime $$\sqrt{(\langle |x(t+\tau)-x(t)|^2\rangle_{th})}\simeq\sqrt{(\frac{\hbar}{2\pi m\Gamma})}\omega_c\tau $$ if $\hbar\omega_c\gg\sqrt{(k_BT\hbar/\tau)}$.  The interesting regime for us is clearly the diffusive one and we would like to illustrate this with an example.
As a numerical illustration we can use a free electron gaz in a metal where the temperature $T$ is replaced by $2/3T_F$ where $T_F$ is the Fermi temperature which is typically  $~10^4$ K (i.e. 2 order of magnitudes more than the room temperature $T$). For example for gold we have $\hbar \Gamma= 65.8$ meV, i.e., $\tau_r=\Gamma^{-1}\simeq 6.2\times10^{-14}$ s, and $E_F=K_B T_F=\frac{(h/\lambda_F)^2}{2m_e}=5.53$ eV, i.e., $T_F=6.42 \times 10^4$ K  and $\lambda_F= 0.55$ nm. Importantly in this model the time $\tau_r$ defines the intrinsic collision time of electrons with the crystal so that we are allowed to write $\omega_c\simeq\Gamma$ (i.e. there is only one time scale here). The condition for being in the diffusive regime reads now $\hbar\omega_c\ll\sqrt{(k_BT\hbar\omega_c)}$, i.e., $\hbar\omega_c\ll k_B T$ in agreement with the physical hypothesis $\hbar\omega_n\ll K_B T$. We are thus in the diffusive regime and we can write for the genuine diffusion constant  
\begin{eqnarray}
D=\frac{2}{3}\frac{k_B T_F}{m\Gamma}=\frac{2}{3}\frac{E_F}{\hbar \Gamma}\frac{\hbar}{2m_e}\nonumber\\
\simeq 112\frac{\hbar}{2m_e}=112 D_Q.\label{diff2}
\end{eqnarray} where we introduced the purely quantum diffusion constant 
\begin{eqnarray}D_Q=\frac{\hbar}{2m_e}
\simeq 5.5\times 10^{-5} m^2s^{-1}
\end{eqnarray} defined by Fenyes and Nelson \cite{Fenyes,Nelson} and advocated by Vigier and de Broglie~\cite{Jalons}. Also, in this regime we have $\frac{\lambda_F}{v_F\tau_r}=\pi\frac{\hbar \Gamma}{E_F}\simeq 0.037$ which means than the typical Fermi wavelength $\lambda_F$ is much smaller than the electron mean free path $v_F\tau_r$ and therefore the plane wave approximation applied during the typical relaxation time $\tau_r$ is good enough (i.e., we are in weak dissipation regime).
\section{Final remarks, and Discussion}
\indent Few remarks are important before to reach our conclusion. First, observe that the mechanism we propose here is fundamentally driven by thermal properties and diffusion mechanism. The results obtained  when the effect of quantum potentials can be neglected is thus very close from the classical or semi-classical diffusion calculations. The success of the procedure relies on the factorization \textit{ansatz}  $\rho_{S+T}(x,\{x_n\},t)\simeq\rho_S(x,t)\rho_T(\{x_n\},t)$  and  $|\psi_{S+T}(x,\{x_n\},t)|^2\simeq|\psi_S(x,t)|^2|\psi_T(\{x_n\},t)|^2$ which is reminiscent of the old molecular chaos axiom. If the bath is in quantum equilibrium, i.e., if $\rho_T(\{x_n\},t)=|\psi_T(\{x_n\},t)|^2$  and also in thermal equilibrium the diffusion process \`a la Langevin  will bring the subsystem S to quantum (and thermal) equilibrium with a typical damping parameter $\Gamma$ and a diffusion constant $D$ given by Eq.~\ref{diff} (e.g., Eq.~\ref{diff2}). This relaxation will be done in agreement with the Fokker-Planck or diffusion equation discussed in \cite{Drezet} (where the same diffusion constant $D=\frac{\langle |x(t)-x(0)|^2\rangle_{th}}{2t}$ was deduced from a Kramers-Moyal  expansion).  Of course, for realistic cases where the quantum potential $Q_{S+T}$ can not be neglected the explicit calculation of the diffusion $D_t$ could be much more involved and actually should be a complex function of time and space. Still, the results obtained here give certainly good order of magnitudes for the diffusion parameter $D$.\\ 
\indent A second important issue concerns the value $D_Q=\hbar/2m$. This quantum diffusion constant was postulated by F\"urth, F\'enyes and Nelson using very different stochastic approaches. If we go back to the original work of F\"urth~\cite{Furth} (see also Bohm \cite{Bohm1957}) based on the  formal analogy between the Schrodinger and diffusion equation (see Feynman and Hibbs~\cite{Feynman} for a discussion in the context of the path integral formalism and specially regarding the use of a pseudo diffusion constant $D'=iD_Q$ for probability amplitudes) we find a very appealing argument for justifying the value $D_Q$. Starting with the Brownian motion law written on the crude form $\delta x^2\simeq 2D t$,  where $\delta x$ is the typical path fluctuating variation along $x$, we get $m\frac{\delta x}{t}\delta x\simeq 2mD$. If we identify $\frac{\delta x}{t}$ with a typical fluctuating velocity variation $\delta v_x$ we get a kind of Heisenberg relation where $2Dm$ plays the role of $\hbar$. The identification $2Dm=\hbar$ leads thus to $D_Q$, i.e., to a purely quantum diffusion constant~\footnote{De Broglie  using a condition of stability on the particle guidance  by the wave obtained in \cite{Jalons} a quite similar result $D=\frac{4\pi}{3} n D_Q$ where $n$ in an integer.}. The reasoning is of course extremely rough since the `velocity' $\frac{\delta x}{t}$ is not  in general identical with the genuine uncertainty $\delta \dot{x}=\delta v_x$ on the velocity $v_x=\dot{x}$. More precisely, Eq.~\ref{einstein}, i.e, the Langevin theory used in the limit $\Gamma t\gg1$ corresponding to time larger than the relaxation time, implies $\delta v_x^2=\langle \dot{x}^2\rangle=D\Gamma(=K_BT /m)$ and $\delta x^2=\langle x^2\rangle=2Dt$  and we thus get 
\begin{eqnarray}
\delta x\delta v_x=\sqrt{(\Gamma t/2)}\frac{\delta x^2}{t}=\sqrt{(\Gamma t/2)}2D\gg2D\label{machen} 
\end{eqnarray}
 Comparing this inequality with the Heisenberg relation $\delta x \delta v_x\gtrsim \hbar/m$ we see  that the identification $D=D_Q$ is still possible if we admit that we are working with semi-classical states for which $\delta x \delta v_x\gg \hbar/m$.  However, if we consider the value Eq.~\ref{diff2} with $D\gg D_Q$  and insert it in Eq.~\ref{machen} we see that this also implies $\delta x \delta v_x\gg \hbar/m$ so that the F\"urth-Bohm intuitive result $D=D_Q$~\cite{Furth,Bohm1957} is not strongly imposed or required by the theory. Actually, we see that it is better to consider $D_Q$ as a standard quantum limit (SQL) in the sense given by Braginsky to this notion in the context of quantum measurement theory~\cite{Braginsky}. Indeed, we know from this theory that the optimum  in precision for measuring the position and momentum of a free particle during a time $t$ are given by $\Delta x_{\textrm{SQL}}\simeq\sqrt{\frac{\hbar t}{2m}} $  and $\Delta p_{\textrm{SQL}}\simeq\sqrt{\frac{\hbar m }{2t}}=m \frac{\Delta x_{\textrm{SQL}}}{t}$. Clearly, here we have a Brownian motion with $D=D_Q$. The meaning of this SQL measurement procedure becomes clear if we remember that decoherence models can interpret the environment (i.e., our thermostat  T) interacting with the particle  of mass $m$ (i.e. our system S) as a form of complex measurement~\cite{zurek,Barnett}. The SQL  value $D=D_Q$ therefore fixes such typical  quantum bound for the interaction with T.\\
\indent It is important also to comment briefly on the difference between our approach and the one followed by Nelson~\cite{Nelson}. Nelson starts from a time symmetric perspective and considers two stochastic evolutions:  forward and backward   associated with respectively  future and past dynamics with respect to a given time $t$.   He proposes (for a single particle) two Brownian equations $d\textbf{x}_{\pm}(t)=(\textbf{u}+\textbf{v})dt+d\textbf{w}_{\pm}(t)$ where  $d\textbf{w}_{\pm}(t)$ is a Wiener process such as the conditional expectation with respect to the present time $t$ reads $E_t[d\textbf{w}_{\pm}(t)\otimes d\textbf{w}_{\pm}(t)]=2D \textbf{I}dt$ (in tensorial notations and using the It\^{o} formalism) with $D$ a diffusion constant which in this approach must be chosen as $D=D_Q$. Here $\textbf{u}(t)=\boldsymbol{\nabla} S/m$ and $\textbf{v}(t)=D\frac{\boldsymbol{\nabla} \rho}\rho$ (where $\rho$ is the density of probability in the configuration space) are called respectively current and osmotic velocities and  in particular $\textbf{u}(t)$ is identical to the one used in the deterministic PWI. Nelson then derives two Fokker-Planck equations (for the forward and Backward motions) and obtains, by addition, the conservation law $\partial_t\rho=-\boldsymbol{\nabla}(\rho\textbf{u})$.  The dynamics of Nelson, which is time symmetric, relies on some assumptions needed to recover the velocities $\textbf{u}(t)$ and $\textbf{v}(t)$ and thus in order to go back to the Schrodinger equation for the wave function  $\psi=\sqrt{\rho} e^{iS}$ (see for example~\cite{Bacciagaluppi,Luis}). The main issue concerns however the extension to the many-body problem and Nelson himself recognized~\cite{Nelson} (see also Cushing~\cite{Cushing}) that his approach, when correctly extended for $N$ particles, leads to some form on nonlocality driven by the stochastic bath.  This nonlocality is actually even stronger than in the PWI since the noise term carries its own nonlocality (added thus to the usual quantum potential). In the present work we followed the deterministic approach of PWI in order to reduce the number of unwanted assumptions (i.e. following a kind of Occam principle) and the nonlocality of the bath is associated with usual quantum entanglement with the environment. This has huge consequences since it means that within the PWI relaxation does not occur all the time (unlike in Nelson's view) but is actually limited to the regime of interacting systems. For example, interacting atoms or electrons  will naturally present such a relaxation but free particles will not (even though entanglement with the bath could be of course preserved after the interaction).  We point out that an alternative approach to Nelson's was later advocated by Bohm and Hiley \cite{Hiley} in which they attempted (following the initial goal of Vigier and Bohm) to derive a stochastic process (different from Nelson's) by adding a Osmotic velocity term $\textbf{v}(t)$ to the PWI with a diffusion constant not necessarily fixed to $D=D_Q$. This approach also leads to a relaxation mechanism where the quantum equilibrium $\rho=|\psi|^2$ appears as an attractor. Interestingly the two models predic  a similar trend to equilibrium\footnote{In \cite{Drezet} we derived the H-theorem starting with $H=\int dx \rho\ln{(f)}$ and with the density of probability $\rho=f|\psi|^2$. We also used two Fokker-Planck equations $\partial_t\rho=-\nabla(\rho v)+D\nabla^2\rho$ and $\partial_t|\psi|^2=-\nabla( |\psi|^2 v)+D\nabla^2(|\psi|^2)$ to obtain the inequality $\frac{d}{dt}H=-\int dxD |\psi|^2(\nabla f)^2/f \leq 0$ which is the H-theorem for our problem~\cite{Drezet}. In Bohm and Hiley work~\cite{Hiley} we have instead with our notations $\partial_t|\psi|^2=-\nabla( |\psi|^2 v)$ and $\partial_t\rho=-\nabla(\rho v)+D\nabla(\rho\nabla\ln{f})$ which lead again to the formula $\frac{d}{dt}H=-\int dx D |\psi|^2(\nabla f)^2/f\leq 0$. Therefore, both methods lead to the same rapid convergence to quantum equilibrium $f=1$.}. We emphasize that both  the Bohm and Hiley and Nelson models suffer from the same arbitrariness and difficulties of interpretation concerning the nonlocality driven by the thermal bath and for these reasons these models are not considered here (while this issue was problematic for Nelson \cite{Nelson}, Bohm and Hiley strongly advocated this nonlocality as a key feature of this stochastic approach: without it it would not be possible to justify the EPR paradox and to obtain a violation of Bell inequalities).\\                      
\indent Finally, a last remark should be done concerning the method used in this work. Indeed, while our work relied on the usual Hamiltonian method of coupling a small system S to  a thermostat T, (i.e., in full agreement with the standard canonical quantization for open systems~\cite{cohen}), this is certainly not the only possible approach. The issue goes back at least to the seminal work by Wigner and Weisskopf~\cite{Wigner} for introducing a complex energy or Hamiltonian in optics~\cite{Landau}. In the same vein a rigorous formalism for non-Hermitian Hamiltonians was used by Dekker~\cite{Dekker} for deriving the Fokker-Planck decoherence/diffusion equation associated with Brownian motion~\cite{Caldeira}. A modified Schrodinger equation including dissipation was proposed by Kostin~\cite{Kostin1972} and is known as the Schrodinger-Langevin equation. In the context of the PWI this approach leads to a pure state description of the particle trajectory since we can define a wave function for the dissipative system without using degrees of freedom for the thermal bath. However, in the Kostin approach, in analogy with Langevin's work we can introduce fluctuational forces associated with a white noise and the approach is thus merely phenomenological (alternative approaches have been proposed by Sanz and coworkers based on the remarkable Caldirola-Kanai formalism for dissipative systems~\cite{Sanz2014b}). In the PWI one can get an  intuitive picture of the Kostin equation starting from the modified Hamilton-Jacobi-Langevin equation~\footnote{Actually we should  replace$\Gamma S(x,t)$ by $\Gamma (S(x,t)-\langle S(t)\rangle)$ if we want to preserve the energy definition $\langle \hat{H}\rangle=\langle[\frac{\hat{p}^2}{2m}+V(\hat{x})-\hat{x}F(t)]\rangle$. } 
\begin{equation}
-\partial_t S(x,t)=\frac{(\nabla S(x,t))^2}{2m}+V(x) +Q(x,t)-xF(t) +\Gamma S(x,t)\label{truec}
\end{equation}             
where $Q(x,t)=-\frac{\hbar^2\Delta a(x,t)}{2ma(x,t)}$ is a quantum potential  and $F(t)$ is a fluctuating force. By taking the gradient and using the guidance postulate $mx\dot{x}=\nabla S$ we immediately get the Langevin equation 
 \begin{eqnarray}
m\ddot{x}(t)=-\nabla(V(x(t))+Q(x(t),t))
-m\Gamma\dot{x}(t)+F(t)	\label{2ff}
\end{eqnarray} which is very similar to Eq.~\ref{2f}. By adding the probability conservation $\partial_t a^2=-\nabla(a^2\nabla S/m)$ and introducing the Kostin wave function $\Psi_K(x,t)=a(x,t)e^{iS(x,t)/\hbar}$ we immediately  deduce the nonlinear Schrodinger-Langevin equation~\footnote{From footnote 3 and \cite{Kostin1972} we emphasize that adding a term  $-\Gamma\langle S(t)\rangle)$ in Eq.~\ref{truec} means adding a term $-\Gamma\Psi_K(x,t)(\int dx' |\Psi_K(x',t)|^2\frac{\hbar}{2i}\Gamma \textrm{ln}[\Psi_K(x',t)/\Psi_K(x',t)])$ in Eq.~\ref{kos}.}
 \begin{eqnarray}
i\hbar\partial_t \Psi_K(x,t)=-\frac{\nabla^2}{2m}\Psi_K(x,t)+[V(x)+Q(x,t)]\Psi_K(x,t)\nonumber\\
-xF(t)\Psi_K(x,t)+\frac{\hbar}{2i}\Gamma \textrm{ln}[\Psi_K(x,t)/\Psi_K(x,t)]\Psi_K(x,t).\label{kos}
\end{eqnarray}   
While this approach (reviewed in a recent book~\cite{BookNassar}) is interesting there are few reasons why we don't consider it here:
First, the theory breaks time symmetry due to the presence of the dissipative term in Eq.~\ref{truec}, also it is as we explained non linear due to the presence of the unusual log term in Eq.~\ref{kos}.  Most importantly, however the model is stochastic due to the presence of the random force $F$ acting as a white noise. This means that the action $S$ as well becomes a stochastic quantity since for every determination of $F$ we have a new solution  for $S$ or $\Psi_K$ (in agreement with the original philosophy of Langevin's model). However, the exact nature of this stochastic space is not clear and the approach is actually more an alternative model like Nelson's stochastic approach was. The connection with the PWI is not clear in particular because it relies also  on the exact conservation of  the probability flow  $\partial_t a^2=-\nabla(a^2\nabla S/m)$ despite the fact that $S$ is fluctuating. In the approach defended here, decoherence and entanglement with the Bath are key and therefore the nature of the stochastic evolution space is clear. In our approach the probability conservation occurs only for the full system S+T and if we average on the degrees of freedom of the bath we get as explained in \cite{Drezet} a Fokker-Planck  or diffusion equation like  $\partial_t a^2\simeq -\nabla(a^2v)+D\nabla^2a^2$ or $\partial_t \rho \simeq -\nabla(\rho v)+D\nabla^2\rho$  (where $\rho(x,t)$ is a reduced probability) which involves the constant $D$ of the Brownian motion driven by the interaction with the bath T. Our approach is intended  for explaining the convergence to quantum equilibrium $\rho\simeq a^2$ and in \cite{Drezet} we showed how diffusion  linked to quantum correlation and entanglement with a thermal bath can lead to this fundamental statistical requirement of the PWI (while the Kostin model, like Nelson's approach, assumes already this postulate). The Langevin equation studies done in the present work not only complete the previous article \cite{Drezet} but also shows how realistic quantum model of the interaction between particles could lead to a realistic picture of relaxation in the PWI. We think that this opens new possibilities for describing non-equilibrium situations in extreme experimental conditions or at the beginning of our Universe.                  
\section{Acknowledgments}
We would like to thank Thomas Durt, Alexandre Matzkin and Christophe Couteau for providing the framework for fruitful discussions concerning Quantum Foundations at Marseille and Troyes in 2016 and 2017.
\appendix
\section{Appendix}
\subsection{About thermal equilibrium in the PWI} 
\indent The non relativistic PWI interpretation is a theory for particles in the configuration space  associated with coordinates $q$ and not a statistical theory in the phase space with canonical coordinates  $q$  and momenta $p$. This has huge consequences since  the basic probability densities are defined as $\rho(q,t)$ and not $\eta(q,p,t)$. Actually, Takabayasi~\cite{Takabayasi,livre,Holland} was the first to point out that  in the PWI we can define densities in the phase space restricted by the Hamilton-Jacobi constraints $p=\nabla S(q,t)$. We have thus in the case of quantum equilibrium $\eta_\psi(q,p,t)=|\psi(q,t)|^2\delta(p-\nabla S(q,t))$ which corresponds to a pure state. However, in order to define a statistical thermal equilibrium for a thermostat we have to introduce a mixture of let say energy states which leads to a reduced density matrix $\hat{\rho}=e^{-\frac{H_{th}}{k_B T}}/Z$  where $H_{th}$ is the bath Hamiltonian and $Z$ the canonical partition  function. This actually means a mixture of wave functions~\cite{Bohmmixture,Hiley} and a  phase space density 
\begin{eqnarray}
      \eta_{th.}(q,p,t)=\sum_E |\psi_E(q,t)|^2\delta(p-\nabla S_E(q,t))\frac{e^{-\frac{E}{k_B T}}}{Z}
\end{eqnarray} This density is not always convenient to use in the PWI for instance when we consider energy average like $\langle E\rangle_{th}=\sum_E E \frac{e^{-\frac{E}{k_B T}}}{Z}$ which in the PWI reads
\begin{eqnarray}
  \langle E\rangle_{th}=\int dq dp \sum_E |\psi_E(q,t)|^2\delta(p-\nabla S_E(q,t)) E\frac{e^{-\frac{E}{k_B T}}}{Z}.
\end{eqnarray} However, since we have $E=\frac{(\nabla S_E(q,t))^2}{2m}+V(q)+Q_E(q)=-\partial_tS_E=H_\psi(x,p,t)$, where the quantum potential $Q_E(q)=-\frac{\hbar^2\Delta|\psi_E(q)|}{2m|\psi_E(q)|}$ is specific of each energy states considered, we can not define a wave-function independent Hamiltonian for the mixture such as $\langle E\rangle_{th}=\int dq dp H(q,p) \eta_{th.}(q,p) $. Therefore, in the PWI the configuration space supersedes the phase space. Still, the concept of mixture in the configuration space is worth and we can safely use 
\begin{eqnarray}
  \langle E\rangle_{th}=\sum_E\int dq \psi_E(q,t)^\ast\hat{H}\psi_E(q,t)\frac{e^{-\frac{E}{k_B T}}}{Z}\nonumber\\
	=\sum_E\int dq  |\psi_E(q,t)|^2[\frac{(\nabla S_E(q))^2}{2m}+V(q)+Q_E(q)] \frac{e^{-\frac{E}{k_B T}}}{Z}.
\end{eqnarray}   
\indent Moreover, the main issue in equilibrium thermodynamics is to obtain this mixture from a pure quantum states. Within the standard  density matrix formalism this is done by taking a huge system and by taking a trace or average over the many degrees of freedom associated with `the rest of the universe'. Physically this means complex interactions and decoherence so as to justify the reduced density matrix $\hat{\rho}_{T}=e^{-\frac{\hat{H}_{th}}{k_B T}}/Z$  from a universal pure state $\hat{\rho}_U=|\Psi_U\rangle\langle \Psi_U|$. This fits quite well with the PWI if we write for any observable $\hat{A}_T$ acting on the thermostat $\langle \hat{A}_T\rangle=\int\int dx_rdx_T \Psi_U(x_r,x_T,t)^\ast\hat{A}_T\Psi_U(x_r,x_T,t)\simeq \sum_E\int dx_T\psi_E(x_T,t)^\ast\hat{A}_T\psi_E(x_T,t)\frac{e^{-\frac{E}{k_B T}}}{Z}$. 
where the label $r$ refers to the rest of the universe degrees of freedom and $\Psi_U(x_r,x_T,t)$ is the universal wave function for the  entangled state involving both the thermostat  T and  the rest of universe r. Moreover, in the PWI the fundamental quantities are the particle trajectories which must be defined from the global wave function  $\Psi_U(x_r,x_T,t)$. The reduced density matrix formalism allows us to define effective paths for the system T after tracing over the degrees of freedom associated with the rest of the universe. For this we define the reduced density matrix as $\hat{\rho}_T=Tr_r[\hat{\rho}_U]$ and  we have
\begin{eqnarray}
\langle x_T|\hat{A}_T\hat{\rho}_T|x'_T\rangle=\int dx_r \Psi_U(x_r,x'_T,t)^\ast\hat{A}_T\Psi_U(x_r,x_T,t)\nonumber\\
\simeq \sum_E \psi_E(x'_T,t)^\ast\hat{A}_T\psi_E(x_T,t)\frac{e^{-\frac{E}{k_B T}}}{Z}
\end{eqnarray} For the probability current operator $\hat{J}_T(x_T)=\frac{|x_T\rangle\langle x_T|\hat{P}_T+\hat{P}_T|x_T \rangle\langle x_T|}{2m}$ we can thus define the effective velocity as  $v_{eff.,T}(x_T,t)=\frac{\langle x_T|\hat{J}_T(x_T)\hat{\rho}_T|x_T\rangle}{\langle x_T|\hat{\rho}_T|x_T\rangle}$, i.e.,
\begin{eqnarray}
v_{eff.,T}(x_T,t)\simeq \frac{\sum_E |\psi_E(x_T)|^2\frac{\nabla_TS_E(x_T)}{m}\frac{e^{-\frac{E}{k_B T}}}{Z}}{\sum_E |\psi_E(x_T)|^2\frac{e^{-\frac{E}{k_B T}}}{Z}}
\end{eqnarray}
This mean Bohmian velocity was advocated in the recent recent years by Appleby~\cite{Appleby} and Sanz \cite{Sanz2014} in the context of decoherence. Alternatively we can take an ensemble point of view and decide to not define this mean velocity. Then by keeping each term of the sum with energy $E$ we  attribute a velocity $\frac{\nabla_TS_E(x_T)}{m}$ to each individual `pure' state in the mixture. This is the strategy used in this work for the thermostat.  
\subsection{Coherent state of the harmonic oscillator and the PWI}
\indent  The usual method for coupling an harmonic oscillator to a thermal bath of oscillators is to suppose that a given time, let say $t=0$, the system S+T is factorisable with a full density matrix $\hat{\rho}=|S\rangle\langle S|\otimes \hat{\rho}_{th.}$ where $|S\rangle$ describes the pure state of the system S while the thermostat  T is characterized by the mixture $\hat{\rho}_{th.}=\otimes_n\hat{\rho}_{th.}^{(n)}$.  For each degrees of freedom of the bath T labeled by $n$ we have $\hat{\rho}_{th.}^{(n)}=\sum_m \frac{e^{-\frac{m\hbar \omega_n}{k_B T}}}{Z_n}|m^{(n)}\rangle\langle m^{(n)}|$ where $|m^{(n)}\rangle$ is a Fock state for the Hilbert space associated with the $n^{th}$ harmonic oscillator of the bath (the partition function reads $Z_n=(1-e^{-\frac{\hbar \omega_n}{k_B T}})^{-1}\simeq\frac{k_B T}{\hbar \omega_n}\gg 1$ in the high temperature limit).\\
\indent However, as explained in the main text the usual Fock states of the harmonic oscillator are not very convenient for the PWI because these are highly non-classical even in the WKB limit corresponding to high quantum number $m\gg 1$. While this doesn't prevent us to use the Langevin equation, here we found it much easier to work with a different representation of the density matrix $\hat{\rho}_{th.}^{(n)}$ namely the one based on the P-representation of Glauber with coherent states $|\alpha\rangle$. There are several reasons motivating this choice. First, coherent states are robust objects which can be easily obtained during a decoherence process involving subsequent baths and interactions~\cite{zurek}. Therefore, they are the most preferred  and natural basis vectors for our reservoir. Second, while for standard quantum  mechanics all the representations of a density matrix are equivalent this is however not the case in the PWI where an ontological level  is introduced in the discussion~\cite{Appleby}. As we will see the coherent states have nice properties which are well suitable for a classical limit description. From now we will remove the label $n$ and consider a generic harmonic oscillator in thermal equilibrium. Using the P-representation of Glauber it is straightforward to write 
\begin{eqnarray}
\hat{\rho}_{th.}=\int \frac{d^2\alpha}{\pi}\rho_{th.}(|\alpha|)|\alpha\rangle\langle \alpha| \label{matrixx}
\end{eqnarray}
 where $\rho_{th.}(|\alpha|)|=e^{-|\alpha|^2/\langle m\rangle_{th.} }/\langle m\rangle_{th.}$  with $\langle m\rangle_{th.}=(e^{\frac{\hbar \omega}{k_B T}}-1)^{-1}\simeq\frac{k_B T}{\hbar \omega}\gg 1$  defines the P-representation of the thermal state in the high temperature limit. If we introduce the polar form $\alpha=|\alpha|e^{i\sigma}$ we have alternatively
\begin{eqnarray}
\hat{\rho}_{th.}=\int_0^{+\infty}\oint d(|\alpha|^2)\frac{d\sigma}{2\pi}\rho_{th.}(|\alpha|)|(|\alpha|e^{i\sigma})\rangle\langle (|\alpha|e^{i\sigma})|\nonumber\\
\simeq \int_0^{+\infty}\hbar\omega d(|\alpha|^2) \frac{e^{-\frac{\hbar \omega|\alpha|^2}{k_B T}}}{k_B T}\oint\frac{d\sigma}{2\pi}|(|\alpha|e^{i\sigma})\rangle\langle ( |\alpha|e^{i\sigma})|
\end{eqnarray}
With this representation we can conveniently  write any average value $\langle\hat{A}\rangle_{th.}=Tr[\hat{\rho}_{th.}\hat{A}]$ associated with the operator $\hat{A}$ acting on the thermal state as
\begin{eqnarray}
\langle\hat{A}\rangle_{th.}\simeq \int_0^{+\infty}\hbar\omega d(|\alpha|^2) \frac{e^{-\frac{\hbar \omega|\alpha|^2}{k_B T}}}{k_B T}\oint\frac{d\sigma}{2\pi}\langle \hat{A}\rangle_\alpha\label{mean}
\end{eqnarray}  with $\langle \hat{A}\rangle_\alpha=\langle \alpha|\hat{A}|\alpha\rangle=\langle ( |\alpha|e^{i\sigma})|\hat{A}|(|\alpha|e^{i\sigma})\rangle$  the average value on the pure coherent state.\\     
\indent For the PWI we need to consider more explicitly the $x$-representation of the coherent state. Also, the time evolution was not considered and  the previous  description corresponds to the density matrix at a origin time $t_0$. The unitary evolution leads to $|\alpha(t)\rangle=U(t,t_0)|\alpha(t_0)\rangle$ where we have $\alpha(t)=\alpha(t_0)e^{-i\omega (t-t_0)}$, $\sigma=\textrm{Arg}[\alpha(t_0)]$. The density matrix at time $t$ is obtained from Eq.~\ref{matrixx} (which represents the state at time $t_0$) by $U(t,t_0)\hat{\rho}_{th.}(t_0)U^{-1}(t,t_0)$. The average value at time $t$  $\langle\hat{A}\rangle_{th.}(t)$ is still given by the integral Eq.~\ref{mean} but with now $\langle \hat{A}(t)\rangle_\alpha=\langle \alpha(t)|\hat{A}(t_0)|\alpha(t)\rangle$ ($\hat{A}(t_0)$ is the Heisneberg representation of the operator at time $t_0$, i.e.,  the Schrodinger representation of this operator).\\
\indent Now, in the x representation the coherent state of the non interacting harmonic oscillator is characterized by a wave function 
\begin{eqnarray}
\langle x|\alpha(t)\rangle=\psi^{(\alpha)}(x,t)=(\frac{m\omega}{\pi\hbar})^{\frac{1}{4}}e^{-\frac{m\omega}{2\hbar}(x-\sqrt{\frac{2\hbar}{m\omega}}\textrm{Re}[\alpha(t)])^2}e^{iS^{(\alpha)}/\hbar}
\end{eqnarray} 
where the phase is 
\begin{eqnarray}
S^{(\alpha)}/\hbar=\sqrt{\frac{2m\omega}{\hbar}}\textrm{Im}[\alpha(t)]x-\frac{\omega}{2}(t-t_0)+\frac{|\alpha(t_0)|^2}{2}\sin{(2\omega (t-t_0)-2\sigma])}.\nonumber\\
\end{eqnarray} Within the PWI the guidance velocity for such a state is:
\begin{eqnarray}
\dot{x}^{(\alpha)}(t)=\frac{\nabla S^{(\alpha)}}{m}=\sqrt{\frac{2\hbar\omega}{m}}\textrm{Im}[\alpha(t)]=-\sqrt{\frac{2\hbar\omega}{m}}|\alpha(t_0)|\sin{(\omega (t-t_0)-\sigma)}\nonumber\\ \label{sol1}
\end{eqnarray}
which by integration leads  to
\begin{eqnarray}
x^{(\alpha)}(t)=\sqrt{\frac{2\hbar}{m\omega}}|\alpha(t_0)|\cos{(\omega (t-t_0)-\sigma)}+u.\label{sol2}
\end{eqnarray} where $u$ is an integration constant which can take any real value. We emphasize that we have $m\dot{x}^{(\alpha)}(t)=\langle \hat{p}\rangle_\alpha(t)$ and $x^{(\alpha)}(t)-u=\langle \hat{x}\rangle_\alpha(t)=\sqrt{\frac{2\hbar}{m\omega}}\textrm{Re}[\alpha(t)]$. Therefore, since $\langle \hat{x}\rangle_\alpha(t)$ is also the trajectory of the wave packet center of mass, $u_0$ is  thus interpreted as a relative coordinate between the Bohmian particle located at $x^{(\alpha)}(t)$ and the center of mass at time $t$.  Importantly Eq.~\ref{sol2}  inserted in Eq.~\ref{mean} with $\hat{A}=\hat{x}$ leads to $\langle\hat{x}\rangle_{th.}=0$  after averaging on the variable  $\sigma$. From the definition of the random force $F'(t)=\sum_n c_n x_n^{(\alpha_n)}(t)$ in Eq.~\ref{forcenew} we thus deduce  $\langle\hat{F'}\rangle_{th.}=0$ as it should be for such a random force.\\ Moreover, with this PWI dynamic we immediately get for the  particle energy $E^{(\alpha)}(t)$
 \begin{eqnarray}
E^{(\alpha)}(t)=-\partial_t S^{(\alpha)}=\hbar\omega|\alpha(t_0)|^2+\frac{\hbar\omega}{2}\nonumber\\+\omega\sqrt{2m\hbar\omega}u|\alpha(t_0)|\cos{(\omega (t-t_0)-\sigma)} \label{energa}
\end{eqnarray}  which is  not a constant of motion (note that by averaging we have $\langle \hat{H}\rangle_\alpha=\int dx|\psi^{(\alpha)}|^2E^{(\alpha)}(t)=\hbar\omega|\alpha(t_0)|^2+\frac{\hbar\omega}{2}$ which is the standard constant of motion value for a coherent state). Furthermore, the quantum potential: $Q^{(\alpha)}(x,t)=-\frac{\hbar^2\Delta |\psi^{(\alpha)}|}{2m|\psi^{(\alpha)}|}$ is 
\begin{eqnarray}
Q^{(\alpha)}(x,t)=\frac{\hbar\omega}{2}-\frac{m\omega^2(x-\langle \hat{x}\rangle_\alpha(t))^2}{2}\label{quantumpot}
\end{eqnarray}
which in agreement with Eqs.~\ref{sol1},\ref{sol2} leads to the Newton-like equation of motion
\begin{eqnarray}
m\ddot{x}^{(\alpha)}(t)=-\nabla[V(x^{(\alpha)}(t))+Q^{(\alpha)}(x^{(\alpha)}(t),t))]\nonumber\\=-m\omega^2\langle \hat{x}\rangle_\alpha(t)=-m\omega^2(x^{(\alpha)}(t)-u).	
\end{eqnarray} We see that the quantum potential provides an additional restoring force modifying the center of application of the Hook law (note that we have indeed $E^{(\alpha)}(t)=\frac{m\omega^2(x^{(\alpha)}(t))^2}{2}+\frac{m(\dot{x}^{(\alpha)}(t))^2}{2}+Q^{(\alpha)}$).\\
\indent The effect of this dynamic is clear when used for calculating mean values in Eq.~\ref{mean}. Starting with the energy  and the value for $\langle \hat{H}\rangle_\alpha$ we get    
\begin{eqnarray}
\langle\hat{H}\rangle_{th.}\simeq \int_0^{+\infty}\hbar\omega d(|\alpha(t_0)|^2) \frac{e^{-\frac{\hbar \omega|\alpha(t_0)|^2}{k_B T}}}{k_B T}[\hbar\omega|\alpha(t_0)|^2+\frac{\hbar\omega}{2}]\nonumber\\
=K_BT+\frac{\hbar\omega}{2}\simeq K_BT
\label{meanbis}
\end{eqnarray} which must be compared to the classical result without the zero point field energy term~\footnote{In classical mechanics we can calculate the phase volume $\delta \Gamma (E)$ between two ellipses of constant energy $E$ and $E+\delta E$ as $\int_{\delta E} dpdq=\delta(\oint p dq)=\delta E /\nu$ where $2\pi\nu=\omega$. This allows us to define the canonical probability in the volume $\delta \Gamma (E)$ as: $\delta P(E)=\frac{\delta E}{\nu}e^{-\frac{E}{k_B T}}/Z=\frac{\delta E}{K_B T}e^{-\frac{E}{k_B T}}$ where we used the partition function $Z=\int_0^{+\infty}\frac{dE}{\nu}e^{-\frac{E}{k_B T}}=\frac{K_B T}{\nu}$.}. We note that we used directly the value of $\langle \hat{H}\rangle_\alpha$. However, if we instead used the expression for $E^{(\alpha)}(t)$ and inverted the integration $\int dx$ and $\oint d\sigma$ in Eq.~\ref{mean} and $\langle \hat{H}\rangle_\alpha=\int dx|\psi^{(\alpha)}|^2E^{(\alpha)}(t)$ we still naturally obtain the same value Eq.\ref{meanbis} since the  $\oint d\sigma\cos{(\omega (t-t_0)-\sigma)}$ term specific of the PWI vanishes. This again stresses the equivalence between standard quantum mechanics and the PWI.\\
Other mean values are particularly important in the present context. First, from Eq.~\ref{sol1} we have 
\begin{eqnarray}
\langle\frac{(\hat{p})^2}{2m}\rangle_\alpha(t)=\int dx^{(\alpha)}(t)|\psi^{(\alpha)}(x^{(\alpha)}(t),t)|^2\frac{m(\dot{x}^{(\alpha)}(t))^2}{2}\nonumber\\
=\hbar\omega|\alpha(t_0)|^2(\sin{(\omega(t-t_0)-\sigma)})^2
\end{eqnarray} which after averaging on the phase $\sigma$ and the amplitude $|\alpha(t_0)|$ leads to the thermal mean value $\langle\frac{(\hat{p})^2}{2m}\rangle_{th.}=\frac{K_BT}{2}$ in agreement with the classical equipartition theorem. A similar calculation can be done for $\langle m\omega^2\frac{(\hat{x})^2}{2}\rangle_\alpha(t)$ which leads to 
\begin{eqnarray}
\langle m\omega^2\frac{(\hat{x})^2}{2}\rangle_\alpha(t)=\int dx^{(\alpha)}(t)|\psi^{(\alpha)}(x^{(\alpha)}(t),t)|^2\frac{m\omega^2(x^{(\alpha)}(t))^2}{2}\nonumber\\
=\frac{\hbar\omega}{4}+\hbar\omega|\alpha(t_0)|^2(\cos{(\omega(t-t_0)-\sigma)})^2\label{energypot}
\end{eqnarray} and again after averaging on the thermal state  $\langle m\omega^2\frac{(\hat{x})^2}{2}\rangle_{th.}=\frac{\hbar\omega}{2}+\frac{K_BT}{2}$. Eq.~\ref{energypot} is important since it shows the presence of a zero point field (zpf) term  which much be included in the energetic balance. Indeed, from  $\langle Q\rangle_\alpha=\hbar\omega/2-\hbar\omega/4$ we  have $\langle \hat{H}\rangle_\alpha=\langle m\omega^2\frac{(\hat{x})^2}{2}\rangle_\alpha(t)+\langle m\omega^2\frac{(\hat{x})^2}{2}\rangle_\alpha(t)+\langle Q\rangle_\alpha=\hbar\omega|\alpha(t_0)|^2+\frac{\hbar\omega}{2}$ in agreement with Eq.~\ref{energa}.\\ \indent The presence of this zpf contribution is important when we calculate the  two-times force correlation $C_F^{(PWI)}(\tau)$ taking into account the bath with the various harmonic oscillators labeled by $n$. We get explicitly   
\begin{eqnarray}
C_F^{(PWI)}(\tau)=\sum_n c_n^2\int_0^{+\infty}\frac{\hbar\omega_n d(|\alpha_n(t_0)|^2) }{k_B T} e^{-\frac{\hbar \omega_n|\alpha_n(t_0)|^2}{k_B T}}\oint \frac{d\sigma_n}{2\pi}I^{(PWI)}_{\alpha_n}(t,\tau)\nonumber\\
\label{meanforce}
\end{eqnarray} with $I^{(PWI)}_{\alpha_n}(t,\tau)=\int dx_n^{(\alpha_n)}(t)|\psi^{(\alpha_n)}(x_n^{(\alpha_n)}(t),t)|^2 x_n^{(\alpha_n)}(t+\tau)x_n^{(\alpha_n)}(t)$. Using the Liouville theorem which allows us to write $$dx_n^{(\alpha_n)}(t)|\psi^{(\alpha_n)}(x_n^{(\alpha_n)}(t),t)|^2=dx_n^{(\alpha_n)}(t_0)|\psi^{(\alpha_n)}(x_n^{(\alpha_n)}(t_0),t_0)|^2$$ and  inserting  Eq.~\ref{sol2} in the definition of $I_{\alpha_n}(t,\tau)$ we get
\begin{eqnarray}
I^{(PWI)}_{\alpha_n}(t,\tau)=\int_{-\infty}^{+\infty}du_n (\frac{m_n\omega_n}{\pi\hbar})^{\frac{1}{2}}e^{-\frac{m_n\omega_n}{2\hbar}u_n^2}[u_n^2\nonumber\\+\frac{2\hbar}{m_n\omega_n}]|\alpha_n(t_0)|^2\cos{(\omega_n(t-t_0)-\sigma_n)}\cos{(\omega_n(t+\tau-t_0)-\sigma_n)}
\label{meanforcebis}
\end{eqnarray} and therefore  
\begin{eqnarray}
\oint\frac{d\sigma_n}{2\pi}I^{(PWI)}_{\alpha_n}(t,\tau)=\frac{\hbar\omega_n}{2m_n\omega_n^2}+\frac{\hbar\omega_n}{m_n\omega_n^2}|\alpha_n(t_0)|^2\cos{(\omega_n\tau)}
\label{meanforcetri}
\end{eqnarray} which implies the force correlation
 \begin{eqnarray}
C_F^{(PWI)}(\tau)=\sum_n \frac{c_n^2}{m_n\omega_n^2}\frac{\hbar\omega_n}{2}+k_BT\sum_n \frac{c_n^2}{m_n\omega_n^2}\cos{(\omega_n\tau)}.
\label{meanforce}
\end{eqnarray} We emphasize that the correlator $C_F^{(PWI)}(\tau)$ used here relies on the definition of $I^{(PWI)}_{\alpha_n}(t,\tau)$ valid in the PWI where deterministic trajectories can be calculated .  In the standard formalism  we instead use the definition  $I^{(Standard)}_{\alpha_n}(t,\tau)=\langle\hat{x}_n(t+\tau)\hat{x}_n(t) \rangle_{\alpha_n}$ which in the Schrodinger picture reads (omitting the n index)  
\begin{eqnarray}
I^{(Standard)}_{\alpha}(t,\tau)=\textrm{Tr}[\hat{\rho}(t_0)\hat{x}(t+\tau)\hat{x}(t)]\nonumber\\=\int dx \int dx' x x'\psi^{\ast,(\alpha)}(x',t+\tau) K(x',t+\tau;x,t) \psi^{(\alpha)}(x,t)
\end{eqnarray} where we inserted the Kernel $K(x',t+\tau;x,t)$ for the Schrodinger equation in the x representation. This formulation was specifically used by Feynman and Hibbs and Feynman and Vernon~\cite{FeynmanHibbs} in the path integral formalism in connections with coupled harmonic oscillators. This distinction is central if one want to interpret properly correlators in various interpretations of quantum mechanics~\cite{tas} and have a self consistent description of quantum measurements. More precisely, a two-times measurements of position at $t_2=t+\tau$ and $t_1=t$ would lead  following Wigner formula~\cite{laloe} to the correlator: 
$\int dx \int dx'x x'\textrm{Tr}[\hat{\rho}(t_0)|x,t_1\rangle\langle x,t_1||x',t_2\rangle\langle x',t_2||x,t_1\rangle\langle x,t_1|]$ which explicitly reads in the Schrodinger picture:
 \begin{eqnarray}
\int dx \int dx'x x'|K(x',t_2;x,t_1)|^2|\psi^{(\alpha)}(x,t_1)|^2.\label{wigner}
\end{eqnarray} This formula differs both from $I^{(PWI)}_{\alpha}(t,\tau)$ and $I^{(Standard)}_{\alpha}(t,\tau)$. Moreover, Eq.~\ref{wigner} can be compared to $I^{(PWI)}_{\alpha}(t,\tau)$ if we write   
    \begin{eqnarray}
I^{(PWI)}_{\alpha}(t,\tau)=\int dx \int dx'x x'P(x',t_2|x,t_1)|\psi^{(\alpha)}(x,t_1)|^2.\label{wignerpwi}
\end{eqnarray} where $P(x',t_2|x,t_1)=\delta(x'-X(t_2|x_1,t_1))$ is the conditional probability for the particle to be located at $x'$ at time $t_2$ knowing that it was located at  $x$ at time $t_1$. Since the evolution is deterministic  in the PWI  the probability is a delta function~\cite{Drezet,Holland} where $X(t_2|x-1,t_1)$ is the `Bohmian' trajectory linking univocally points $x$ and $x'$ at their respective times $t_1$ and $t_2$. $P(x',t_2|x,t_1)$ is in general clearly different from $|K(x',t_2;x,t_1)|^2$ because as stated before the PWI deals with hidden variables having an existence independently  of measurements and we didn't speak about measurements in the present article. Naturally, if we introduce a  two-times measurement  then Eq.~\ref{wigner} will ultimately become the good formula to use and the PWI will agree with that providing we introduce correctly the measurement protocol with a wave function `collapse' at time $t_1$.      


\end{document}